# The Gaussian Many-to-One Interference Channel with Confidential Messages

Xiang He   Aylin Yener
Wireless Communications and Networking Laboratory
Electrical Engineering Department
The Pennsylvania State University, University Park, PA 16802
xxh119@psu.edu   yener@ee.psu.edu

*Abstract*—We investigate the $K$-user many-to-one interference channel with confidential messages in which the $K$th user experiences interference from all other $K-1$ users, and is at the same time treated as an eavesdropper to all the messages of these users. We derive achievable rates and an upper bound on the sum rate for this channel and show that the gap between the achievable sum rate and its upper bound is $\log_2(K-1)$ bits per channel use under very strong interference, when the interfering users have equal power constraints and interfering link channel gains. The main contributions of this work are: (i) nested lattice codes are shown to provide secrecy when interference is present, (ii) a secrecy sum rate upper bound is found for strong interference regime and (iii) it is proved that under very strong interference and a symmetric setting, the gap between the achievable sum rate and the upper bound is constant with respect to transmission powers.

## I. INTRODUCTION

In a wireless environment, interference is always present. Traditionally, interference is viewed as a harmful physical phenomenon that should be avoided. Yet, from the secrecy perspective, if interference is more harmful to an eavesdropper, it can be a resource to protect confidential messages. To fully appreciate and evaluate the potential benefit of interference on secrecy, the fundamental model to study is the interference channel with confidential messages. Indeed, this model with two users has been studied extensively up to date, e.g., [1]–[6].

The case with more than two users, by comparison, is not well explored. Difficulties in solving the K-user case, $K \geq 3$, exist in both the achievability and the converse. For achievability, there is no known scheme for the strong interference regime. The strong interference scenario is usually dismissed for the two-user case since the achievable secrecy rates are much smaller than those achievable under weak interference regime [2]. In contrast, the K-user strong interference case is quite different, because the $K-1$ interfering users can in fact protect each other in the strong interference regime and a substantial amount of secrecy rate can be achieved. The conventional wisdom says in strong interference, the receiver should remove the interference before decoding the intended message. Yet, in secrecy problems, we have to face the question on how to remove the interference when the receiver is *not supposed to decode* the interference. This problem is addressed in [7] for the case where all links are i.i.d. fading under a continuous distribution, and interference alignment in temporal domain leads to achievable rates. Yet, if the channel is static, this method is not applicable and new methods are needed.

In this aspect, progress in interference channel without secrecy constraint points to the use of lattice codes, which is essentially interference alignment in signal space. This approach allows decoding the sum of interference without knowing each component in it. Notable results include [8] where lattice codes are used for interference alignment for a many-to-one Gaussian interference channel. The same idea also applies to a fully connected interference channel [9], [10].

In this work, we focus on the Gaussian many-to-one interference channel first studied in [8], in an effort to investigate the effects of interference in the context of secrecy. We use lattice codes to achieve secrecy for this model and use the tool first introduced in [11] which computes secrecy rates when the lattice code has a nested structure [12]. Notably, the structure of the lattice we use differs from that used in interference channels without secrecy constraints [8]–[10], and accordingly so does its error probability analysis [12].

For the converse, known results are limited to the case where the eavesdropper observes a weaker channel than the legitimate receiver [3], [4]. The upper bound from [1] is general, yet is difficult to evaluate for the Gaussian case due to the presence of the auxiliary random variables. While the upper bound in [2] is applicable to the strong interference case, it is shown therein to be quite loose for strong interference, mainly because the genie information used in deriving the upper bound provides too much information to the legitimate receiver. Another contribution of this work is providing a good sum rate upper bound for the many-to-one interference channel under strong interference.

Under very strong interference, we show that the gap between our upper bound and our achievable sum rate is $\log_2(K-1)$ bits under certain uniform interference conditions. We observe that in this setting, for fixed transmission power $P$, the cost of secrecy constraints per user diminishes when the number of users $K \to \infty$. This means that as the number of users gets large, the secrecy constraints induce a negligible rate penalty for each user, i.e., secrecy comes for free.

The following notation is used throughout the paper: $C(x) = 0.5\log_2(1+x)$. $A_{1,...,K}$ represents the set $\{A_1, A_2, ..., A_K\}$. $\oplus \sum_{i=1}^n A_i$ is used as a shorthand for

$A_1 \oplus A_2 ... \oplus A_n$, and $R_{sum}$ for $\sum_{i=1}^{K} R_i$.

## II. PRELIMINARIES

In this section, we provide the preliminaries related to nested lattice codes, which will be useful in providing the achievable rates in Section IV. Let $\Lambda$ denote a lattice in $\mathbf{R}^N$ [12], i.e., a set of points which is a group closed under real vector addition. The modulus operation $x \mod \Lambda$ is defined as $x \mod \Lambda = x - \arg\min_{y \in \Lambda} d(x,y)$, where $d(x,y)$ is the Euclidean distance between $x$ and $y$. The fundamental region of a lattice $\mathcal{V}(\Lambda)$ is defined as the set $\{x : x \mod \Lambda = x\}$.

Let $t_1, t_2, ..., t_K$ be $K$ numbers taken from $\mathcal{V}(\Lambda)$. Then, we have the following *representation theorem*:

*Theorem 1:* $\sum_{k=1}^{K} t_k$ is uniquely determined by $\{T, \sum_{k=1}^{K} t_k \mod \Lambda\}$, where $T$ is an integer such that $1 \leq T \leq K^N$.

*Remark 1:* The theorem is a purely algebraic result and does not rely on the statistics of $t_{1,...K}$. The case with $K = 2$ was proved in [11]. The proof here is similar and is hence omitted due to the space limit. For $K = 2$, theorem 1 implies that modulus operation looses at most one bit per dimension of information if $t_1, t_2 \in \mathcal{V}$.

## III. SYSTEM MODEL

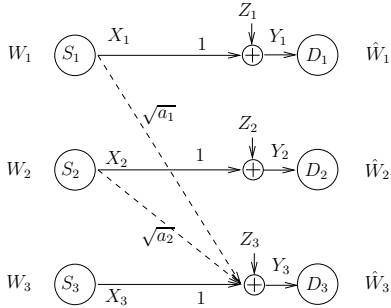

Fig. 1. Many-to-one Gaussian interference channel. number of users K=3

We consider the many-to-one Gaussian interference channel [8] in Figure 1. The average power constraint for node $S_i$ is $P_i$. $Z_i, i = 1, ..., K$ are independent Gaussian random variables with zero mean and unit variance. The channel gain of the link between $S_i$ and $D_i$ is unity. The channel gain between $S_i$ and $D_K$ is $\sqrt{a_i}$.

Node $S_i$ sends a message $W_i$ to node $D_i$, while keeping it secret from the other receivers. Hence, for $W_{1,...,K-1}$, node $D_K$ is viewed as an eavesdropper. Let the signal received by $D_K$ over $n$ channel uses be $Y_K^n$. The corresponding secrecy constraint is given by:

$$\lim_{n \to \infty} \frac{1}{n} H\left(W_{1,2,...,K-1} | Y_K^N\right) = \lim_{n \to \infty} \frac{1}{n} H\left(W_{1,2,...,K-1}\right) \tag{1}$$

## IV. ACHIEVABLE RATES

Without loss of generality, we assume there is a $j$, such that

$$a_j P_j \leq a_i P_i, \quad \forall i \tag{2}$$

*Theorem 2:* Let $K \geq 3$. Define $P_{min} = \min\{P_1, ..., P_{K-1}\}$. If

$$a_j > \max\left\{\left(\frac{P_K + 1}{P_j}\right)\left(\frac{K-2}{K-1} + P_{min}\right), \frac{P_K + 1}{P_j}\right\} \tag{3}$$

Then the following sum secrecy rate is achievable

$$R_{sum} = [(K-2)R_{min} - \log_2(K-1)]^+ + R_K \tag{4}$$

where $R_{min} = C(P_{min}), R_K = C(P_K)$.

*Proof:* Let $(\Lambda, \Lambda_c)$ denote a nested lattice structure in $\mathbf{R}^N$, where $\Lambda_c$ is the coarse lattice.

Node $S_i, i = 1, ..., K$ constructs its input to the channel over $N$ channel uses, $X_i^N$, as follows: The code book has rate $R_i$ and is composed of points $t_i \in \Lambda_i \cap \mathcal{V}(\Lambda_{c,i})$. The first $K - 1$ users use the same lattice. Hence we require $R_i \equiv R$, $\Lambda_i \equiv \Lambda, \Lambda_{c,i} \equiv \Lambda_c$ for $i = 1...K - 1$. Let $d_i$ be the dithering noise, which is uniformly distributed over $\mathcal{V}(\Lambda_{c,i})$. We assume the lattice is scaled properly such that

$$\frac{1}{N \int_{x \in \mathcal{V}(\Lambda_{c,i})} dx} \int_{x \in \mathcal{V}(\Lambda_{c,i})} \|d_i\|^2 \, dx = 1 \tag{5}$$

Let $P = a_j P_j$, where $j$ is defined in (2). Define $x \oplus y$ as $x \oplus y = (x + y) \mod \Lambda_c$. Further, define $U_i^N$ and $X_i^N$ as:

$$U_i^N = t_i^N \oplus d_i^N, i = 1, ..., K-1 \tag{6}$$

$$U_K^N = (t_K^N + d_K^N) \mod \Lambda_{c,K} \tag{7}$$

$$X_i^N = \frac{\sqrt{P}}{\sqrt{a_i}} U_i^N, i = 1, ..., K-1, \quad X_K^N = \sqrt{P_K} U_K^N \tag{8}$$

In order for $D_i, i = 1, ..., K-1$ to correctly decode $t_i$, based on [12, Theorem 5], the probability of decoding error will go to zero as $N \to \infty$, if

$$R \leq C(P_i), i = 1, ..., K-1 \tag{9}$$

The signal received by $D_K$ over $N$ channel uses is given by

$$Y_K^N = \sqrt{P}(\sum_{i=1}^{K-1} U_i^N) + \sqrt{P_K} U_K^N + Z_K^N \tag{10}$$

Node $D_K$ decodes the interference first: It selects a constant $\alpha$ and computes $\hat{Y}_K^N$ as shown below [12]: Let $\gamma = \sqrt{P_K/P}$. Let $Z_K'^N = Z_K^N/\sqrt{P}$.

$$\hat{Y}_K^N = (\frac{\alpha}{\sqrt{P}} Y_K^N - \sum_{i=1}^{K-1} d_i^N) \mod \Lambda_c \tag{11}$$

$$= (\alpha(\sum_{i=1}^{K-1} U_i^N + \gamma U_K^N + Z_K'^N) - \sum_{i=1}^{K-1} d_i^N) \mod \Lambda_c \tag{12}$$

$$= (\sum_{i=1}^{K-1} t_i^N + (\alpha - 1)\sum_{i=1}^{K-1} U_i^N + \alpha(\gamma U_K^N + Z_K'^N))) \mod \Lambda_c \tag{13}$$

$\alpha$ is chosen so that the variance of the effective noise term

$$Z_{eff}^N = (\alpha - 1)\left(\sum_{i=1}^{K-1} U_i^N\right) + \alpha(\gamma U_K^N + Z_K'^N) \quad (14)$$

per dimension is minimized. Under the optimal $\alpha$, the effective noise variance is $\frac{P_X P_N}{P_X + P_N}$, where $P_X = \gamma^2 + \frac{1}{P} = \frac{P_K+1}{P}$. $P_N = K - 1$.

Clearly the effective noise $Z_{eff}^N$ is not Gaussian. However, $U_i^N$ can be approximated with a Gaussian distribution as shown below [12, (200)]:

$$f_{U_i^N}(x) \leq e^{N\varepsilon(\Lambda_{c,i})} f_{O_i^N}(x) \quad (15)$$

where $O_i, i = 1, ..., K, \sim \mathcal{N}(0, \sigma_i^2 I)$, where $\sigma_i^2$ is the average power per dimension of a random variable uniformly distributed over the smallest ball covering $\mathcal{V}(\Lambda_{c,i})$. $\varepsilon(\Lambda_{c,i})$ is defined as [12, (67)]:

$$\varepsilon(\Lambda_{c,i}) = \log\left(\frac{R_{u,i}}{R_{l,i}}\right) + \frac{1}{2}\log 2\pi e G_N^* + \frac{1}{N} \quad (16)$$

where $R_{u,i}, R_{l,i}$ are the covering radius and effective radius of $\Lambda_{c,i}$ respectively. $G_N^*$ is the normalized average power of $N$-sphere and converges to $\frac{1}{2\pi e}$ as $N \to \infty$. The lattice is designed to be good for covering. Hence $\frac{R_{u,i}}{R_{l,i}} \to 1$ as $N \to \infty$. $\sigma_i^2$ is bounded below [12, Lemma 6][1]:

$$\frac{N}{N+2} \leq \sigma_i^2 \leq \left(\frac{R_{u,i}}{R_{l,i}}\right)^2 \quad (17)$$

Note that this approximation property in (15) is invariant under scaling. This means for any $c > 0$, we have:

$$f_{cU_i^N}(x^N) \leq e^{N\varepsilon(\Lambda_{c,i})} f_{cO_i^N}(x^N) \quad (18)$$

In addition, for any two independent random variables $U_1^N, U_2^N$ that have the approximation property given by (15), the probability density distribution of their sum can be approximated as

$$f_{U_1^N + U_2^N}(x^N) \leq e^{N\varepsilon(\Lambda_{c,1}) + N\varepsilon(\Lambda_{c,2})} f_{O_1^N + O_2^N}(x^N) \quad (19)$$

Define $\tilde{Z}^N$ as

$$\tilde{Z}^N = (1-\alpha)\left(\sum_{i=1}^{K-1} O_i^N\right) + \alpha(\gamma O_K^N + Z_K'^N) \quad (20)$$

Based on the two properties described above, we find the effective noise can be approximated by $\tilde{Z}^N$ as follows:

$$f_{Z_{eff}^N}(x) \leq e^{(K-1)N\varepsilon(\Lambda_c) + N\varepsilon(\Lambda_{c,K})} f_{\tilde{Z}^N}(x) \quad (21)$$

Node $D_K$ attempts to decode $\oplus \sum_{i=1}^{K-1} t_i$. The approximation in (21) enables us to apply the analysis in [12, Theorem 5], that the probability of decoding error will go to 0 as $N \to \infty$ when

$$R \leq 0.5 \log_2\left(\frac{1}{\frac{P_X P_N}{P_X + P_N}}\right) = 0.5 \log_2\left(\frac{1}{K-1} + \frac{P}{P_K + 1}\right) \quad (22)$$

[1]The $\frac{P_X P_N}{P_X+P_N}$ in [12, Lemma 6] corresponds to the average power per dimension of $U_i$ here.

After subtracting the interference, the remainder of the interference signal is

$$\gamma U_K^N \oplus Z_K'^N \quad (23)$$

We next show if

$$P_K + 1 < P \quad (24)$$

then this signal can be approximated by

$$\gamma U_K^N + Z_K'^N \quad (25)$$

by which we mean:

$$\lim_{N \to \infty} \Pr((\gamma U_K^N \oplus Z_K'^N) \neq (\gamma U_K^N + Z_K'^N)) = 0 \quad (26)$$

As $N \to \infty$, $\gamma U_K^N + Z_K'^N$ can be approximated by $\gamma O_K^N + Z_K'^N$, such that

$$\Pr\left(\gamma U_K^N + Z_K'^N \notin \mathcal{V}(\Lambda_c)\right) \leq e^{N\varepsilon(\Lambda_{c,K})} \Pr\left(\gamma O_K^N + Z_K'^N \notin \mathcal{V}(\Lambda_c)\right) \quad (27)$$

Let $\mu = \frac{1}{\gamma^2 + 1/P} = \frac{P}{P_K + 1}$. Because the shaping lattice is Poltyrev-good [12], if $\mu > 1$, we have

$$\Pr\left(\gamma O_K^N + Z_K'^N \notin \mathcal{V}(\Lambda_c)\right) \leq e^{-N(E_P(\mu) - o_N(1))} \quad (28)$$

where $E_P(\mu)$ is the Poltyrev exponent defined in [12, (56)]. Since $E_p(\mu)$ is positive for $\mu > 1$, we have the approximation given in (25). Node $D_K$ then tries to decode $t_K$ from (25). Based on [12, Theorem 5], the probability of decoding error will go to zero as $N \to \infty$, if

$$R_K < C(P_K) \quad (29)$$

In summary, there are three types of error events at the destination:

1) $E_1$: $D_K$ incorrectly decodes the modulus sum of the interference.
2) $E_2$: $E_1$ does not occur; and (25) does not equal (23).
3) $E_3$: $E_1, E_2$ do not occur; and $D_K$ incorrectly decodes the lattice point $t_K^N$ after subtracting the interference.

If (22), (24), (29) hold, then

$$\lim_{N \to \infty} \Pr\left(\bigcup_{i=1}^{3} E_i\right) = \lim_{N \to \infty} \sum_{i=1}^{3} \Pr(E_i) = 0 \quad (30)$$

Also (9) must be met in order for $t_i$ to be correctly decoded at $D_i, i = 1, ..., K$.

We next bound the mutual information leaked to the eavesdropper as follows.

$$H(t_{1,...,K-1}^N | Y_K^N, d_i^N, i = 1, .., K) \quad (31)$$
$$\geq H(t_{1,...,K-1}^N | Y_K^N, X_K^N, Z_K^N, d_i^N, i = 1, .., K) \quad (32)$$
$$= H(t_{1,...,K-1}^N | \sum_{i=1}^{K-1} U_i^N, d_i^N, i = 1, .., K) \quad (33)$$

Let $T$ is the integer in Theorem 1, which is used to recover $\sum_{i=1}^{K-1} U_i^N$ from $\oplus \sum_{i=1}^{K-1} U_i^N$. $1 \leq T \leq (K-1)^N$. Then (33) becomes:

$$H(t_{1,...,K-1}^N | \oplus \sum_{i=1}^{K-1} U_i^N, T, d_i^N, i=1,..,K) \quad (34)$$

$$= H(t_{1,...,K-1}^N | \oplus \sum_{i=1}^{K-1} t_i^N, T) \quad (35)$$

$$\geq H(t_{1,...,K-1}^N | \oplus \sum_{i=1}^{K-1} t_i^N) - H(T) \quad (36)$$

The first term in (36) can be bounded as follows:

$$H(t_{1,...,K-1}^N | \oplus \sum_{i=1}^{K-1} t_i^N) = \sum_{j=1}^{K-1} H(t_j^N | t_{1,...,j-1}^N, \oplus \sum_{i=1}^{K-1} t_i^N) \quad (37)$$

$$= \sum_{j=1}^{K-1} H(t_j^N | \oplus \sum_{i=j}^{K-1} t_i^N) = \sum_{j=1}^{K-2} H(t_j^N) = (K-2)NR \quad (38)$$

Hence the mutual information leaked to the eavesdropper is bounded as: $I(t_{1,...,K-1}^N; Y_K^N, d_i^N, i=1,...,K) \leq N(R + \log_2(K-1))$

With this preparation, we can now derive the secrecy rate. We notice that when (9), (22), (24), (29) hold, node $D_K$ can decode the modulus sum of the interference, and then decode $t_K$. Hence the channel can be viewed as composed of two parts: one part is a direct link from $S_K$ to $D_K$. The other part is the orthogonal MAC wire-tap channel considered in [4], where the main channel is composed of $K-1$ orthogonal components, and the eavesdropper observes a MAC channel. The signal received by the eavesdropper is the interference received by $D_K$. The difference is that this MAC wire-tap channel has discrete inputs $t_1^N, ..., t_{K-1}^N$. Each channel use in this new channel corresponds to $N$ channel uses of the original channel. Following a similar argument in [13], for this equivalent channel, the following secrecy rate $(R_{1,e}, ..., R_{K-1,e})$ is achievable:

$$0 \leq R_{i,e} \leq H(t_i^N) - R_{i,x}, \ R_{i,x} \geq 0, i=1,...,K-1 \quad (39)$$

$$\sum_{i=1}^{K-1} R_{i,x} = I(t_{1,...,K-1}^N; Y_K^N, d_i^N, i=1,...,K) \quad (40)$$

Finally, it can also be verified that (3) holds, (24) is fulfilled and (22) is looser than (9) and hence becomes redundant. Under (3), $R = C(P_{min}), i = 1, ..., K - 1$. The result in the theorem follows by choosing $R_{i,x}$ as

$$R_{i,x} = \frac{N}{K-1}(C(P_{min}) + \log_2(K-1)) \quad (41)$$

∎

*Remark 2:* When using nested lattice codes to this interference channel, we had to overcome two difficulties: (1) The error probability analysis in [12] requires the noise to be Gaussian, while in an interference channel, the interference plus noise is in general non-Gaussian. We managed to get around this via the property that a good lattice code, after dithering, "looks like" Gaussian noise [12]. (2) In the decoder of a nested lattice code, a nonlinear modulus operation [12] must be applied to the received signal. This operation causes distortion to the signal even after the decoded part of the signal is subtracted out and renders the use of layered encoding and decoding in [10] not straightforward. This is resolved by proving that the probability of having distortion in fact goes to 0 as $N \to \infty$.

## V. UPPER BOUND ON THE SECRECY SUM RATE

Assume $a_i \geq 1, i = 1...K - 1$. Let $n$ be the total number of channel uses. Define $V^n$ as: $V^n = \sum_{i=1}^{K-1} \sqrt{a_i} X_i^n + Z_K^n$. Then we have the following lemma:

*Lemma 1:*

$$nR_{sum} \leq I(W_{1,...,K-1}; Y_{1,...,K-1}^n) - I(W_{1,...,K-1}; V^n) + I(X_K^n; Y_K^n | X_{1,...,K-1}^n) + n\varepsilon \quad (42)$$

where $\lim_{n \to \infty} \varepsilon = 0$.

*Proof Outline:* The two user case ($K = 2$) has been proved in [3, Appendix]. The same technique is used here to prove Lemma 1. The derivation starts from [3, (41)], $W_1$ being replaced by $W_{1,...,K-1}$, $Y_1$ being replaced by $Y_{1,...,K-1}$, $X_1$ being replaced by $X_{1,...,K-1}$, $Y_2$ being replaced by $Y_K$. The $V_1^n$ therein is replaced by $V^n$. Then, we can prove

$$nR_{sum} - n\varepsilon \leq I(W_{1,...,K-1}; Y_{1,...,K-1}^n) - I(W_{1,...,K-1}; V^n) + I(W_{1,...,K-1}; V^n | Y_K^n) + I(X_K^n; Y_K^n) \quad (43)$$

It can then be shown, following a similar derivation to [3, Appendix (46)-(57)], that

$$I(W_{1,...,K-1}; V^n | Y_K^n) + I(X_K^n; Y_K^n) \leq I(X_K^n; Y_K^n | X_{1,...,K-1}^n) \quad (44)$$

Hence we have (42). ∎

Let $\tilde{V}^n = \sum_{i=1}^{K-1} \sqrt{\frac{a_i}{c}} X_i^n + \sqrt{\frac{1}{c}} Z_K^n + \sqrt{1 - \frac{1}{c}} \tilde{Z}_K^n$, where $c = \max\{a_i, i = 1, ..., K - 1\}$. $\tilde{Z}_K^n$ is a length-$n$ vector that has the same distribution as $Z_K^n$ but is independent from $Z_K^n$. Then we have the following lemma:

*Lemma 2:*

$$R_{sum} \leq \lim_{n \to \infty} \frac{1}{n} (\sum_{i=1}^{K-1} I(X_i^n; Y_i^n) - I(X_{1,...,K-1}^n; \tilde{V}^n)) + \lim_{n \to \infty} \frac{1}{n} I(X_K^n; Y_K^n | X_{1,...,K-1}^n) \quad (45)$$

*Proof Outline:* Because $\tilde{V}^n$ is a degraded version of $V^n$, from Lemma 1 and data processing inequality, we have

$$nR_{sum} \leq I(W_{1,...,K-1}; Y_{1,...,K-1}^n) - I(W_{1,...,K-1}; \tilde{V}^n) + I(X_K^n; Y_K^n | X_{1,...,K-1}^n) + n\varepsilon \quad (46)$$

where $\lim_{n \to \infty} \varepsilon = 0$. Next, we extend the derivation in [4, (58),(65)-(68)] to the first two terms, by replacing $Y^n$ with

$Y^n_{1,...,K-1}$. The derivation in [4, (58),(65)-(68)] corresponds to the case of $K-1=2$ here. It is important to note that $\tilde{V}^n$ is not the signal received by the eavesdropper. Hence the channel is not equivalent to the channel considered in [4], which has different secrecy constraints. However, as we have shown above, the derivation in [4, (58),(65)-(68)] does not invoke any secrecy constraint. Hence these steps can still be applied here and we have the lemma. ∎

*Theorem 3:* When $a_i \geq 1, i=1...K-1$, the sum secrecy rate is upper bounded by

$$R_{sum} \leq \sum_{i=1}^{K} C(P_i) - C\left(\frac{\sum_{i=1}^{K-1} a_i P_i}{(K-1)c}\right) \quad (47)$$

where $c = \max\{a_i, i=1...K-1\}$.

*Proof Outline:* The theorem follows by evaluating the bound in Lemma 2. This is done by extending [4, Theorem 4]. [4, Theorem 4] corresponds to the case with $K-1=3$.

Let $h_i = a_i/c, i=1,...,K-1$. Then it can be shown that the first limit in (45) is upper bounded by

$$\sum_{i=1}^{K-1} C(P_i) + C\left(\frac{\sum_{i=1}^{K-1} h_i P_i}{K-1}\right) \quad (48)$$

The main technique is the generalized entropy power inequality [14]. Since no secrecy constraint is invoked in its derivation, its result is still applicable here. This, along with the fact that $I(X_K^n; Y_K^n | X_{1,...,K-1}^n) \leq nC(P_K)$, gives us the result in the theorem. ∎

## VI. COMPARISON OF THE ACHIEVABLE RATE AND THE UPPER BOUND

When $a_i = a, P_i = P_{min}, i=1...K-1$, and the condition on $a$ given by (3) is fulfilled, the achievable secrecy sum rate, given by Theorem 2, becomes

$$R^a_{sum} = [(K-2)C(P_{min}) - \log_2(K-1)]^+ + C(P_K) \quad (49)$$

The upper bound on the secrecy sum rate, given by Theorem 3 becomes

$$R^{ub}_{sum} = (K-2)C(P_{min}) + C(P_K) \quad (50)$$

It is easy to see that the gap between upper bound and lower bound is at most $\log_2(K-1)$ bits per channel use.

The cost in rate, paid by first each $K-1$ users, following from (41), is $\frac{1}{K-1}(C(P_{min}) + \log_2(K-1))$. We see that, for fixed $P_{min}$, this rate loss goes to 0 as $K \to \infty$. This observation is demonstrated in Figure 2.

## VII. CONCLUSION

In this work, we have derived achievable secrecy rates for K ($K \geq 3$) user Gaussian many-to-one interference channel, and an upper bound on the secrecy sum rate. The achievability technique is general and applies to the full connected $K$-user interference channel as well [15]. The converse utilizes a combination of techniques in [3], [4]. Although both techniques were designed for weak interference, we show their combination provides a good sum rate upper bound for the strong interference case.

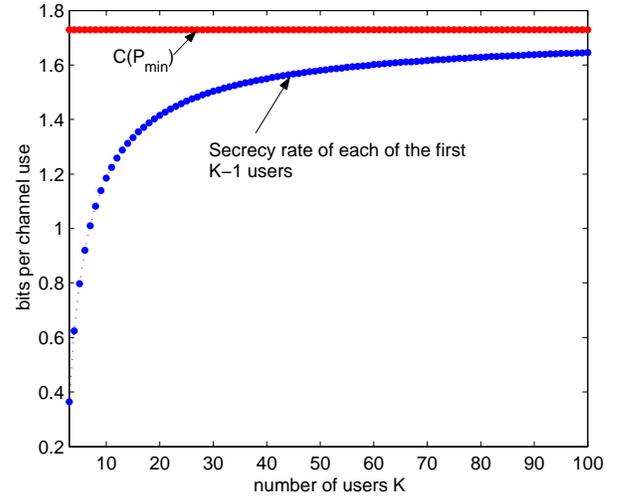

Fig. 2. Rate penalty paid for secrecy per user reduces as the number of users $K$ increases. $P_{min} = 10$.